\documentclass{elsart}
\usepackage[english]{babel}
\usepackage{bbm}
\usepackage{mathrsfs}
\usepackage{pifont}
\usepackage{bbding}
\usepackage{}
\usepackage{stmaryrd}
\usepackage{amsmath}
\usepackage{amscd}

\usepackage{graphicx}
\usepackage{subfigure}
\usepackage{amssymb}
\usepackage[left,pagewise]{lineno}
\usepackage{extarrows}
\usepackage{floatrow}
\newfloatcommand{capbtabbox}{table}[][\FBwidth]
\newtheorem{Th}{\bf Theorem}[section]
\newtheorem{Cor}[Th]{\bf Corollary}
\theoremstyle{definition} 
\theoremstyle{remark}
\newtheorem{Def}[Th]{\bf Definition}
\newtheorem{Pro}[Th]{\bf Proposition}
\newtheorem{Exam}[Th]{\bf Example}

\newtheorem{Rem}[Th]{\bf Remark}
\newtheorem{Lem}[Th]{\bf Lemma}
\newtheorem{Fa}[Th]{\bf Fact}

\newcommand{\down}{\mathord{\downarrow}\hspace{0.05em}}
\newcommand{\up}{\mathord{\uparrow}\hspace{0.05em}}
\newcommand{\uuar}{\mathord{\Downarrow}\hspace{0.05em}}
\newcommand{\dda}{\mathord{\Uparrow}\hspace{0.05em}}

\newcommand{\p}[1]{{\bf Proof.} #1 \ $\Box$}

\begin{document}
\begin{frontmatter}
\title{Some problems about co-consonance of topological spaces}
\author{Zhengmao He, Bin Zhao}
\thanks{This work is supported by the National Natural Science Foundation of
China (Grant no. 11531009).}
\thanks{Corresponding author: B. Zhao.} \thanks{E-mail addresses:
hezhengmao@snnu.edu.cn,  zhaobin@snnu.edu.cn.}

\address{School of Mathematics and Statistics, Shaanxi Normal University,\\ Xi'an 710119, P.R. China} \maketitle

\begin{abstract}\quad In this paper, we first prove that the retract of a consonant space (or co-consonant space) is consonant (co-consonant). Using this result, some related results have obtained. Simultaneously, we proved that (1) the co-consonance of the Smyth powerspace $P_{S}(X)$ implies the co-consonance of $X$ under a necessary condition; (2) the co-consonance of $X$ implies the co-consonance of the smyth powerspace under some conditions; (3) if the lower powerspace $P_{H}(X)$ is co-consonant, then $X$ is co-consonant; (4) the co-consonance of $X$ implies the co-consoance of the lower powerspace $P_{H}(X)$ with some sufficient conditions.
\end{abstract}

\begin{keyword} retraction ; consonant space; co-consonant space; Smyth powerspace; lower powerspace
\vspace*{0.6cm}


\end{keyword}

\end{frontmatter}
\vspace{0.01cm}
\section{Introduction}
\setlength{\parskip}{0.5\baselineskip}
\addtolength{\parskip}{4pt}
\par\quad Given a topological space $X$, there are two topologies on $\mathcal{O}(X)$, let $\mathcal{O}(X)$ be the poset of all open sets of $X$ with inclusion order. In this paper, we consider three kinds of topologies on $\mathcal{O}(X)$. One is the topology $\tau$ that has a base $\{\square Q\mid Q\ \mbox{is  compact in}\  X \}$, where $\square Q=\{V\in\mathcal{O}(X)\mid Q\subseteq V\}$, the second is the Scott topology $\sigma(\mathcal{O}(X))$ and the third is the upper topology $\upsilon(\mathcal{O}(X))$. It is not difficult to find that $\tau\subseteq\sigma(\mathcal{O}(X))$. A topological space $X$ is consonant if $\tau=\sigma(\mathcal{O}(X))$. Equivalently, a space $X$ is consonant if the Upper Kuratowski topology and the co-compact on the set of all closed sets of $X$ are equal (see \cite{RS17}). Consonant spaces have been researched by many scholars and the more conclusions about consonance can see \cite{MA61,MLR19,RS17}. Given a poset $P$, the upper topology $\upsilon(P)$ on $P$ is a topology that has a subbase $\{P\setminus\down x\mid x\in P\}$. Clearly, $\upsilon(P)\subseteq\sigma(P)$, where $\sigma(P)$ is the Scott topology on $P$. A topological space is co-consonant if the upper topology and the Scott topology on the open set lattice $\mathcal{O}(X)$ coincide (see \cite{WR30}). Compared with the consonance, the co-consonance is lack of research. In this paper, we focus our interest on co-consonant spaces.

\quad Retract is an ordinary relation of topological spaces. There are many topological properties which are preserved by retraction, such as sobriety, well-filteredness, d-spaces. Naturally, it arises the following two questions:

${\rm (1)} $ Is the retract of a consonant space consonant?

${\rm (2)} $ Is the retract of a co-consonant space co-consonant?

 In section 3, we give a positive answer. Using this result, some related conclusions are discovered.

\quad Given a topological space $X$, there are two constructions of topological spaces, namely, the lower powerspace $P_{H}(X)$ and smyth powerspace $P_{S}(X)$ (see \cite{GG03,JG13}). Recently, M. Brecht and T. Kawai proved that $X$ is consonant iff the lower powerspace and Smyth powerspace on $X$ is commute (see \cite{WR30}). Furthermore, M. Brecht and T. Kawai asked that whether the consonance is preserved by the Smyth powerspace construction. This question is first answered by Z. Lyu, Y. Chen, and X. Jia (see \cite{AS35} ). It has proved in \cite{AS35}  that if $X$ is consonant and for natural number $n$,
$$\Sigma(\prod\limits^{n}\mathcal{O}(X))=\prod\limits^{n}\Sigma(\mathcal{O}(X)),$$
then $P_{S}(X)$ is consonant. Similarly, Y. Chen, H. Kou and Z. Lyu also showed that if $X$ is consonant and for natural number $n$,
$$\Sigma(\prod\limits^{n}\mathcal{O}(X))=\prod\limits^{n}\Sigma(\mathcal{O}(X)),$$
 then $P_{H}(X)$ is consonant (see \cite{GHY30}). Naturally, it arises the following questions.

{\rm $(3)$} Is $X$ co-consonant when $P_{s}(X)$ is co-consonant?

{\rm $(4)$} Is the smyth powerspace $P_{s}(X)$ of a co-consonant space $X$ co-consonant?

{\rm $(5)$} Is the lower powerspace $P_{H}(X)$ of a co-consonant space $X$ co-consonant?

{\rm $(6)$} Is $X$ co-consonant when $P_{H}(X)$ is co-consonant?

In section 4, we will give some answers for these questions.

\section{Preliminaries}
\quad Given a poset $P$ and $A\subseteq P$, let $$\up A=\{x\in P\mid a\leq x\ \mbox{for some}\ a\in A\}$$ and $$\down A=\{x\in P\mid x\leq a\ \mbox{for some}\ a\in A\}.$$
For every $x\in P$, we write $\down x$ for $\down\{x\}$  and $\up x$ for $\up\{x\}$ .

\quad Let $P$ be a poset and $x, y\in P$. We say that $x$ is {\em way below} $y$, in symbols $x\ll y$, iff for all directed subsets $D\subseteq P$ for which $\bigvee D$ exists, $y\leq\bigvee D$ implies $x\leq d$ for some $d\in D$.\ If  $\uuar a=\{b\in P\mid b\ll a\}$ is  directed  and $\bigvee \uuar a=a$\ for all \ $a\in P$, we call $P$  a continuous poset. A complete lattice $L$ is called a continuous lattice if $L$ is a continuous poset.

\quad Let $P$ be a poset. A subset $U$ of $P$ is {\em Scott open} (see \cite{GG03})
 if (i) $U$$=\up U$ and (ii) for any directed subset $D$, $\bigvee D\in U$ implies
$D\cap U\neq \emptyset$ whenever $\bigvee D$ exists. The Scott open sets on $P$ form the Scott topology $\sigma(P)$. The Scott space $(P, \sigma(P))$ will be simply written as $\Sigma(P)$. The upper topology $\upsilon(P)$ on poset $P$ is a topology that has $\{P\setminus\down x\mid x\in P\}$ as a subbase.

\quad For a $T_{0}$ space $(X,\tau)$, the {\em specialization order} $\leq$ on $X$ is defined by $x\leq y$ if and only if $x\in cl(\{y\})$ (see \cite{GG03}, p. 42]). We use $\mathcal{O}(X)$ ($\Gamma(X)$) to denote the lattice of all open(closed) subsets of $X$. Note that for each subset $A\subseteq X$, $\up A$ is equal to the intersection of all the open sets containing $A$ and we say $\up A$ is the saturation of $A$. A subset $A\subseteq X$ is saturated if $A=\up A$.

\quad For a topological space $X$, we shall use $\mathcal{Q}(X)$ to denote the poset of all nonempty
compact saturated subsets of $X$ with the reverse inclusion order. The upper Vietoris topology on $\mathcal{Q}(X)$ is the topology that has $\{\Box U\mid U\in\mathcal{O}(X)\}$ as a base, where $\Box U=\{K\in\mathcal{Q}(X)\mid K\subseteq U\}$. The upper Vietoris topological space (or called smyth powerspace) is denoted by $P_{S}(X)$. Naturally, $\{\diamond U\mid U\in\Gamma(X)\}$ is a subbase of the closed sets of $P_{S}(X)$, where $\diamond U=\{Q\in\mathcal{Q}(X)\mid Q\cap U\neq\emptyset\}$. Then the specialization order of the upper space $P_{S}(X)$ is the reverse inclusion order.

\quad Given a topological space $X$, the lower powerspace on $\Gamma(X)$ is the topology that has $\{\diamond U\mid U\in\mathcal{O}(X)\}$ as a subbase, where $\diamond U=\{V\in\Gamma(X)\mid U\cap V\neq\emptyset\}$. The lower powerspace is denoted by $P_{H}(X)$. Then $\{\Box U\mid U\in\Gamma(X)\}$ is a subbase of the closed sets of $P_{H}(X)$, where $\Box U=\{F\in\Gamma(X)\mid F\subseteq U\}$. One can see that $\diamond (U\cup V)=(\diamond U)\cup(\diamond V)$ for each pair of open sets $U$ and $V$.

\quad Let $X$ be a $T_{0}$ space. A nonempty subset $F$ of $X$ is {\em irreducible}, if for any $A,B\in\Gamma(X)$, $F\subseteq A\cup B$ implies $F\subseteq A$ or $F\subseteq B$. For every $x\in X$,\ $cl(\{x\})$ is an irreducible closed set of  $X$. A topological space $X$ is called to be {\em sober} if every irreducible closed set of $X$ is the closure of an unique singleton set. A topological space $X$ is {\em well-filtered} if for each filtered family $\mathcal{F}$ of compact saturated subsets of $X$ and each open set $U$ of $X$, $\bigcap \mathcal{F}\subseteq U$ implies $F\subseteq U$ for some $F\in\mathcal{F}$.  It is well known that
every sober space is well-filtered (see \cite{GG03}). A topological space $X$ is {\em coherent} if the intersection of two compact saturated subsets is compact. For every complete lattice $L$, the Scott space $\Sigma (L)$ is well-filtered and coherent (see \cite{JAL16, XL17}).

\quad A topological space $X$ is said to be {\em locally compact} if $\forall\ x\in U\in \mathcal{O}(X)$, there is an open set $V$ and a compact subset $K$ such that $x\in V\subseteq K\subseteq U$. A topological space $X$ is said to be  {\em core-compact } if  $(\mathcal{O}(X), \subseteq)$ is a continuous lattice. It has known that every locally compact space is core-compact.

\begin{Lem}(see \cite{GG03}) Let $P$ be a dcpo. Then for each dcpo $Q$, $\sigma(P\times Q)=\mathcal{O}(\Sigma P\times\Sigma Q)$ iff $\Sigma P$ is core-compact.
\end{Lem}

Interestingly, J.Goubault-Larrecq find the following results in his blog which is due to Matthew de Brecht.

\begin{Lem}(see \cite{AU41}) Let $P, Q$ be posets. If $\Sigma P$ and $\Sigma Q$ are first countable, then $\sigma(P\times Q)=\mathcal{O}(\Sigma P\times\Sigma Q)$.
\end{Lem}

\section{The retract of consonant and co-consonant}

In the following, we will give positive answers to the question 1 and question 2 proposed in introduction.

\begin{Def} A $T_{0}$ space $X$ is consonant if for every $\mathcal{F}\in\sigma(\mathcal{O}(X))$ and $U\in\mathcal{F}$, there is $Q\in\mathcal{Q}(X)$ such that $U\in\square Q\subseteq \mathcal{F}$, where $\square Q=\{V\in\mathcal{O}(X)\mid Q\subseteq V\}$.
\end{Def}

\begin{Exam} {\rm (1)} Every locally $k_{\omega}$ space is consonant (see \cite{RS17}).

{\rm (2)}A topological space $X$ is locally compact iff $X$ is a core-compact and consonant space (see \cite{GHY30}).

{\rm (3)} The Sorgenfrey line is not consonant (see \cite{MA61}).
\end{Exam}

\begin{Def}(see \cite{WR30})  A $T_{0}$ space $X$ is co-consonant if for every $\mathcal{F}\in\sigma(\mathcal{O}(X))$ and $U\in\mathcal{F}$, there is a finite subset $\mathcal{E}\subseteq\Gamma(X)$ such that $U\in\bigcap\{\lozenge A\mid A\in\mathcal{E}\}\subseteq \mathcal{F}$, where $\lozenge A=\{V\in\mathcal{O}(X)\mid A\cap V\neq\emptyset\}$.
\end{Def}

\quad Clearly, $X$ is co-consonant iff the upper topology is  agree with the Scott topology on $\mathcal{O}(X)$.

 \begin{Exam} {\rm (1)} If $X$ is a quasi-polish space, then $P_{S}(X)$ is a co-consonant space (see \cite{WR30}).

{\rm (2)} Let $L$ be the Isbell complate lattice (see \cite{KJ82}). We define $\mathcal{O}(\hat{L})=\{\hat{L}\setminus K \mid K\in\mathcal{Q}(L)\}$, where $\hat{L}=L\setminus\{1_{L}\}$ and $1_L$ is the top element of $L$. Then $(\hat{L},\mathcal{O}(\hat{L}))$ is a topological space and $\Sigma(\mathcal{O}(\hat{L}))$ is non-sober (see \cite{HP35}). By Proposition 2.9 in \cite{DU13}, the upper topology on $\mathcal{O}(\hat{L})$ is sober. Hence we can assert that $\sigma(\mathcal{O}(\hat{L}))\neq\upsilon((\mathcal{O}(\hat{L})))$. This implies that $(\hat{L},\mathcal{O}(\hat{L}))$ is not co-consonant.
\end{Exam}

\quad The Example 3.4(2) illustrates that the following conclusion holds. 

\begin{Fa} Let $X$ be a topological space. If $X$ is co-consonant, then $\Sigma(\mathcal{O}(X))$ is sober .
\end{Fa}

Next, we will give two kinds of order spaces which are co-consonant. Given a poset $P$, the Alexandroff topology $\alpha(P)$ is the topology consisting
of all its upper subsets of $P$.

\begin{Pro} Let $P$ be a poset. Then the  Alexandroff topological space $(P,\alpha(P))$ is co-consonant.
\end{Pro}
\p{It is clear that $\upsilon(\alpha(P))\subseteq\sigma(\alpha(P))$. Let $\mathcal{V}\in\sigma(\alpha(P))$ and $V\in\mathcal{V}$ with $V\neq\emptyset$. Note that $V=\bigcup\{\up F\mid F\subseteq V\ \mbox{is a finite non-empty subaset}\}$. It follows from $\mathcal{V}\in\sigma(\alpha(P))$ that there exists a subset $F_{0}=\{x_{1},x_{2},\cdot\cdot\cdot,x_{n}\}$ of $V$ such that $\up F_{0}\in\mathcal{V}$. Clearly $V\in\bigcap\limits_{1\leq m\leq n}(\alpha(P)\setminus\down(P\setminus(\down x_{m})))$. Let $B\in\bigcap\limits_{1\leq m\leq n}(\alpha(P)\setminus\down(P\setminus(\down x_{m})))$. Then $B\in\alpha(P)$ and $B\cap\down x_{m}\neq\emptyset$ for all $1\leq m\leq n$. This means that $x_{m}\in B$ for all $1\leq m\leq n$. Since $\up F_{0}=\bigcup\limits_{1\leq m\leq n}\up x_{m}\in\mathcal{V}$, we have $B\in\mathcal{V}$. So $V\in\bigcap\limits_{1\leq m\leq n}(\alpha(P)\setminus\down(P\setminus(\down x_{m})))\subseteq\mathcal{V}$ and thus $\upsilon(\alpha(P))=\sigma(\alpha(P))$.}

\begin{Pro} Let $P$ be a continuous poset. Then $\Sigma(P)$ is co-consonant.
\end{Pro}
\p{Let $\mathcal{U}\in\sigma(\sigma(P))$ and $U\in\mathcal{U}$. Since $P$ is a continuous poset, we have $U=\bigcup\limits_{u\in U}\dda u$. Note that $S=\{\bigcup\limits_{1\leq s\leq m}\dda u_{s}\mid u_{s}\in U, m\in \mathbb{N}\}$ is directed and $U=\bigcup S$. As $\mathcal{U}\in\sigma(\sigma(P))$, there are finitely $u_{1},u_{2},\cdot\cdot\cdot u_{n}\in U$ such that $\bigcup\limits_{1\leq k\leq n}\dda u_{k}\in\mathcal{U}$. Define a map $f_{n}:\prod\limits^{n}P\longrightarrow \Sigma(\sigma(P))$ as follows $$\forall(x_{1},x_{2},\cdot\cdot\cdot,x_{n})\in\prod\limits^{n}P, f((x_{1},x_{2},\cdot\cdot\cdot,x_{n}))=\bigcup\limits_{1\leq k\leq n}\dda x_{k}.$$
Clearly, $\down u_{1}\times\down u_{2}\times\cdot\cdot\cdot\times\down u_{n}\subseteq f_{n}^{-1}(\mathcal{U})$. Let $V\in\bigcap\limits_{1\leq k\leq n}\diamondsuit \down u_{k}$. Then for all $1\leq k\leq n$, $V\cap \down u_{k}\neq\emptyset$. This means that $u_{k}\in V$. Since $\bigcup\limits_{1\leq k\leq n}\dda v_{k}\subseteq V$ and $\mathcal{U}\in\sigma(\sigma(P))$, we have $V\in\mathcal{U}$. This implies that $U\in\bigcap\limits_{1\leq k\leq n}\diamondsuit \down u_{k}\subseteq\mathcal{U}$. Thus, $\Sigma(P)$ is co-consonant.}

\quad A topological space $X$ is a {\em retract} of a topological space $Y$ if there are two continuous maps $f:X\longrightarrow Y$ and $g:Y\longrightarrow X$ such that $g\circ f=id_{X}$.

\begin{Th} Let $X$ be a $T_{0}$ space. If $X$ is a retract of a consonant space $Y$, then $X$ is consonant.
\end{Th}

\p{ Let $\mathcal{F}\in\sigma(\mathcal{O}(X))$ and $U\in\mathcal{F}$. Since $X$ is a retract of $Y$, there are two continuous maps $f:X\longrightarrow Y$ and $g:Y\longrightarrow X$ such that $g\circ f=id_{X}$. Define a mapping $\alpha:\Sigma(\mathcal{O}(Y))\longrightarrow\Sigma(\mathcal{O}(X))$ as follows:
 $$\forall\ U\in\mathcal{O}(Y)),\alpha(U)=f^{-1}(U).$$
Clearly, $\alpha$ is continuous. So we have $\widetilde{\mathcal{F}}=\alpha^{-1}(\mathcal{F})\in\sigma(\mathcal{O}(Y))$. Since $g\circ f=id_{X}$, $g$ is a surjection and for every subset $A\subseteq X$, $A=f^{-1}(g^{-1}(A))$.
So $f^{-1}(g^{-1}(U))=U$ and thus $g^{-1}(U)\in\widetilde{\mathcal{F}}$. Since $Y$ is consonant and $g^{-1}(U)\in\widetilde{\mathcal{F}}$, there is $Q\in\mathcal{Q}(Y)$ such that $g^{-1}(U)\subseteq\square Q\subseteq\widetilde{\mathcal{F}}$. Note that for each subset $B\subseteq X$, $g(g^{-1}(B))=B$ since $g$ is a surjection.

$\mathbf{Claim}$: $U\in \square\up g(Q)\subseteq\mathcal{F}$.

Clearly, $\up g(Q)\in\mathcal{Q}(X)$ and $\up g(Q)\subseteq\up g(g^{-1}(U))= U$. Whence, $U\in \square\up g(Q)$. For each open set $E\in\square\up g(Q)$, $g(Q)\subseteq\up g(Q)\subseteq E$. This implies that
 $$Q\subseteq g^{-1}(g(Q))\subseteq g^{-1}(E).$$ Therefore, $g^{-1}(E)\in\square Q$ and hence $g^{-1}(E)\in\widetilde{\mathcal{F}}$. This implies that $E=f^{-1}(g^{-1}(E))\in\mathcal{F}$. This means that $\square\up g(Q)\subseteq\mathcal{F}$. Hence, $X$ is consonant.}

\begin{Th} Let $X$ be a $T_{0}$ space. If $X$ is a retract of a co-consonant space $Y$, then $X$ is co-consonant.
\end{Th}

\p{ Let $\mathcal{F}\in\sigma(\mathcal{O}(X))$ and $U\in\mathcal{F}$. Since $X$ is a retract of $Y$, there are two continuous mappings $f:X\longrightarrow Y$ and $g:Y\longrightarrow X$ such that $g\circ f=id_{X}$. Similarly, by the proof of Theorem 3.8, $\widetilde{\mathcal{F}}=\{V\in\mathcal{O}(Y)\mid f^{-1}(V)\in\mathcal{F}\}\in\sigma(\mathcal{O}(Y))$ and $g^{-1}(U)\in\widetilde{\mathcal{F}}$. By the co-consonance of $Y$, there is a finite subset $\mathcal{E}\subseteq\Gamma(Y)$ such that $g^{-1}(U)\in\bigcap\{\lozenge A\mid A\in\mathcal{E}\}\subseteq\widetilde{\mathcal{F}}$. Let $\widetilde{\mathcal{E}}=\{f^{-1}(A)\mid A\in\mathcal{E}\}$. It follows form the continuity of $f$ that $\widetilde{\mathcal{E}}$ is a finite subset of $\Gamma(X)$.

$\mathbf{Claim}$: $U\in\bigcap\{\lozenge B\mid B\in\widetilde{\mathcal{E}}\}\subseteq\mathcal{F}$.

Since $g^{-1}(U)\in\bigcap\{\lozenge A\mid A\in\mathcal{E}\}$, $g^{-1}(U)\cap A\neq\emptyset$ for each $A\in\mathcal{E}$. Take $y\in g^{-1}(U)\cap A$. So $g(y)\in U$ and thus $y=f(g(y))\in A$. Equivalently, $g(y)\in U\cap f^{-1}(A)$. This means that $U\in\bigcap\{\lozenge B\mid B\in\widetilde{\mathcal{E}}\}$. For each $W\in\bigcap\{\lozenge B\mid B\in\widetilde{\mathcal{E}}\}$, $W\cap f^{-1}(A)\neq\emptyset$ for each $A\in\mathcal{E}$.
Take $x\in W\cap f^{-1}(A)$. One can easily check that $f(x)\in g^{-1}(W)\cap A$. This implies that $g^{-1}(W)\in\bigcap\{\lozenge A\mid A\in\mathcal{E}\}$. So $W=f^{-1}(g^{-1}(W))\in\mathcal{F}$ and thus $\bigcap\{\lozenge B\mid B\in\widetilde{\mathcal{E}}\}\subseteq\mathcal{F}$. Therefore $X$ is co-consonant.}

\quad Next, we present some applications of Theorem 3.8 and Theorem 3.9.

\begin{Rem}{\rm(1)} For a complete lattice $L$, $\Sigma L$ is a retract of $\Sigma(\Gamma(\Sigma L))$ (see \cite{PJ81}).

{\rm (2)} Let $X$ be a $T_{0}$ space. Define mappings  $\phi:P_{S}(X)\longrightarrow P_{S}(P_{S}(X))$ and $\varphi:P_{S}(P_{S}(X))\longrightarrow P_{S}(X)$ as follows: $$\forall\ Q\in\mathcal{Q}(X),\phi(Q)=\up_{P_{S}(X)}\xi(Q)$$ and
$$ \forall\ \mathcal{A}\in\mathcal{Q}(P_{S}(X)),\varphi(\mathcal{A})=\bigcup\mathcal{A},$$ where the mapping $\xi : X\longrightarrow P_{S}(X)$ is defined by $\xi(x)=\up x$. Then $\phi$ and $\varphi$ are continuous such that $\varphi\circ\phi=id_{P_{S}(X)}$ and $\phi\circ\varphi\geq id_{P_{S}(P_{S}(X))}$. So we can conclude that $P_{S}(X)$ is a strong retract of $P_{S}(P_{S}(X))$.
\end{Rem}

\begin{Cor}{\rm(1)} Let $L$ be a complete lattice. If $\Sigma(\Gamma(L))$ is consonant (co-consonant), $\Sigma L$ is consonant (co-consonant).

{\rm (2)} Let $X$ be a $T_{0}$ space. If $P_{S}(P_{S}(X))$ is consonant (co-consonant), $P_{S}(X)$ is consonant (co-consonant).
\end{Cor}

\begin{Lem}(see \cite{XXZ20}) Let $L$ be a complete lattice. Then $\mathcal{Q}(L)=\mathcal{Q}(\Sigma L)$ is a complete Hetying algebra.

\end{Lem}

\begin{Rem} Let $L$ be a complete lattice. Then for any $\{Q_{i}\mid i\in I\}\subseteq\mathcal{Q}(L)$, $\bigvee_{i\in I}Q_{i}=\bigcap_{i\in I}Q_{i}$.

\end{Rem}

\begin{Lem} Let $L$ be a complete lattice. If $\Sigma L$ is consonant, then $\Sigma(\mathcal{Q}(L))$ is a retract of $\Sigma(\sigma(\sigma(L)))$.

\end{Lem}

\p{ Define mappings $f:\Sigma(\mathcal{Q}(L))\longrightarrow\Sigma(\sigma(\sigma(L)))$ and $g:\Sigma(\sigma(\sigma(L)))\longrightarrow\Sigma(\mathcal{Q}(L))$ as follows:
$$\forall\ Q\in\mathcal{Q}(L),\  f(Q)=\square Q,$$
and

$$\forall\ \mathcal{F}\in\sigma(\sigma(L)),\  g(\mathcal{F})=\bigcap\mathcal{F}.$$

 Since $\Sigma L$ is consonant and $\mathcal{F}\in\sigma(\sigma(L))$, there exists $\mathcal{K}\subseteq\mathcal{Q}(L)$ such that $\mathcal{F}=\bigcup\{\square K\mid K\in\mathcal{K}\}$. Then $\bigcap\mathcal{F}=\bigcap(\bigcup\{\square K\mid K\in\mathcal{K}\})=\bigcap\limits_{K\in\mathcal{K}}(\bigcap\square K)=\bigcap\mathcal{K}$. By Remark 3.12, we can see $\bigcap\mathcal{F}\in\mathcal{Q}(L)$ and $g$ is well-defined. Clearly, $g\circ f=id_{\Sigma(\mathcal{Q}(L))}$.

Claim: $f$ and $g$ are continuous mappings.

Let $\{Q_{i}\mid i\in I\}\subseteq\mathcal{Q}(L)$ be a directed subset. By Remark 3.12, $\bigvee\{Q_{i}\mid i\in I\}=\bigcap\{Q_{i}\mid i\in I\}$. Then $$f(\bigvee\{Q_{i}\mid i\in I\})=f(\bigcap\{Q_{i}\mid i\in I\})=\square(\bigcap\{Q_{i}\mid i\in I\}).$$
Clearly, $$\bigvee\{f(Q_{i})\mid i\in I\}=\bigcup\{\square Q_{i}\mid i\in I\}\subseteq f(\bigvee\{Q_{i}\mid i\in I\})=\square(\bigcap\{Q_{i}\mid i\in I\}).$$ For each $A\in\square(\bigcap\{Q_{i}\mid i\in I\})$, equivalently $\bigcap\{Q_{i}\mid i\in I\}\subseteq A$. Since $\Sigma L$ is well-filtered, there exists $Q_{i_{0}}$ such that $Q_{i_{0}}\subseteq A$. So $A\in\square Q_{i_{0}}$ and thus $$f(\bigvee\{Q_{i}\mid i\in I\})\subseteq\bigvee\{f(Q_{i})\mid i\in I\}.$$ Hence, $f(\bigvee\{Q_{i}\mid i\in I\})=\bigvee\{f(Q_{i})\mid i\in I\}$. This means that $f$ is Scott continuous. Let $\{\mathcal{F}_{i}\mid i\in I\}\subseteq\sigma(\sigma(L))$ be a directed subset. Then $g(\bigvee\{\mathcal{F}_{i}\mid i\in I\})=g(\bigcup\{\mathcal{F}_{i}\mid i\in I\})=\bigcap(\bigcup\{\mathcal{F}_{i}\mid i\in I\})=\bigcap\{\bigcap\mathcal{F}_{i}\mid i\in I\}=\bigvee\{g(F_{i})\mid i\in I\}$. So $g$ is also Scott continuous. Hence, $\Sigma(\mathcal{Q}(L))$ is a retract of $\Sigma(\sigma(\sigma(L)))$.}

\quad By Theorem 3.8 and Theorem 3.9, we have the following corollary immediately.

\begin{Cor} Let $L$ be a complete lattice. If $\Sigma L$ is consonant, then the following statements hold:

{\rm(1)} if $\Sigma(\sigma(\sigma(L)))$ is consonant, then $\Sigma(\mathcal{Q}(L))$ is consonant;

{\rm(2)} if $\Sigma(\sigma(\sigma(L)))$ is co-consonant, then $\Sigma(\mathcal{Q}(L))$ is co-consonant.
\end{Cor}

\begin{Pro} Let $X$ and $Y$ be a  pair of $T_{0}$ spaces. If $X\times Y$ is co-consonant (consonant), then $X$ is co-consonant (consonant).
\end{Pro}

\p{ Fix a $y\mathcal{}_{0}\in Y$, the mapping $\alpha_{y_{0}}:X\longrightarrow X\times Y$ is given by
$$\forall\ x\in X,\ \alpha_{y_{0}}(x)=(x,y_{0}).$$
For each $A\in\mathcal{O}(X\times Y)$, $A=\bigcup\limits_{i\in I}(U_{i}\times V_{i})$ for some $\{U_{i}\mid i\in I\}\subseteq\mathcal{O}(X)$ and $\{V_{i}\mid i\in I\}\subseteq\mathcal{O}(Y)$.
One can check that
$$ \ \alpha^{-1}_{y_{0}}(A)=\left\{
             \begin{array}{ll}
              \bigcup\limits_{i\in I_{0}}U_{i}, &\ \ I_{0}=\{i\in I\mid y_{0}\in V_{i}\}\neq\emptyset, \\
               \emptyset, &\ \ y_{0}\not\in\bigcup\limits_{i\in I} V_{i}.
             \end{array}
           \right.$$
This means that $\alpha_{y_{0}}$ is a continuous mapping. Clearly, $p\circ\alpha_{y_{0}}=id_{X}$, where $p:X\times Y\longrightarrow X$ is the projection. So we can assert that $X$ is a retract of $X\times Y$. By Theorem 3.8 and Theorem 3.9, $X$ is co-consonant (consonant).}
\section{Co-consonance of powerspaces}

In this section, we will give some answers for those four questions which is proposed in the introduction. Specifically, the answer of question 3 is negative and the partial answers of question 4 and question 5 is given. Furthermore, the answer of question 6 is positive.

\quad Before answering question 3, it is necessary to introduce the concept of strongly compact subsets. A subset $K$ of a topological space $X$ is strongly compact if for each open set $U$, $K\subseteq U$ implies that there is a finite subset $F\subseteq X$ such that $K\subseteq\up F\subseteq U$ (see \cite{WR30}).

\begin{Lem} (see \cite{WR30}) Let $X$ be a co-consonant $T_{0}$ space. Then every compact subset of $X$ is strongly compact.

\end{Lem}

\quad In the following, we prove that the converse of Lemma 4.1 is true if $P_{S}(X)$ is co-consonant.

\begin{Th} Let $X$ be a $T_{0}$ space.  If $P_{S}(X)$ is co-consonant and every compact subset of $X$ is strongly compact, then $X$ is co-consonant.
\end{Th}

\p{Define a mapping $\xi:X\longrightarrow P_{S}(X)$ by $\xi(x)=\up x$, for all $x\in X$. Then $\xi$ is continuous. So the mapping $f:\mathcal{O}(P_{S}(X))\longrightarrow\mathcal{O}(X)$ is well defined, where $f(U)=\xi^{-1}(U)$ for all $U\in\mathcal{O}(P_{S}(X))$. Clearly, $f$ is Scott continuous. For each $\mathcal{H}\in\sigma(\mathcal{O}(X))$ and $U\in\mathcal{H}$. Then $\Box U\in f^{-1}(\mathcal{H})$ and

 \begin{center}$\begin{array}{lll}
f^{-1}(\mathcal{H})&=&\{\bigcup\limits_{i\in I}\Box U_{i}\mid f(\bigcup\limits_{i\in I}\Box U_{i})\in\mathcal{H}\}\\
&=&\{\bigcup\limits_{i\in I}\Box U_{i}\mid \{x\mid \up x\in\bigcup\limits_{i\in I}\Box U_{i}\}\in\mathcal{H}\}\\
&=&\{\bigcup\limits_{i\in I}\Box U_{i}\mid \{x\mid x\in\bigcup\limits_{i\in I}U_{i}\}\in\mathcal{H}\}\\
&=&\{\bigcup\limits_{i\in I}\Box U_{i}\mid\bigcup\limits_{i\in I}U_{i}\in\mathcal{H}\}.\\
\end{array}$\end{center}

 By the continuity of $f$, $f^{-1}(\mathcal{H})\in\sigma(\mathcal{O}(P_{S}(X)))$. Since $P_{S}(X)$ is co-consonant,
there is a finite subset $\mathcal{F}\subseteq\Gamma(P_{S}(X))$ such that $\Box U\in\bigcap\{\lozenge F\mid F\in\mathcal{F}\}\subseteq f^{-1}(\mathcal{H})$. For each $F\in\mathcal{F}$, let $F=\bigcap\{\diamond V_{i}\mid i\in I_{F}\}$, where $\{V_{i}\mid i\in I_{F}\}\subseteq\Gamma(X)$. Since $\Box U\in\lozenge F$ for each $F\in\mathcal{F}$, there exists a $Q_{F}\in\mathcal{Q}(X)$ such that $Q_{F}\subseteq U$ and $Q_{ F}\in F$. By assumption, $Q$ is strongly compact. Then there is a finite subset $N_{F}$ such that $Q_{F}\subseteq\up N_{F}\subseteq U$. Let $\widetilde{\mathcal{F}}=\{cl(\{x\})\mid x\in\bigcup\limits_{F\in\mathcal{F}}N_{F}\}$. Clearly, $\widetilde{\mathcal{F}}$ is a finite subset of $\Gamma(X)$.

$\mathbf{claim}$ 1: $U\in\bigcap\{\lozenge A\mid A\in\widetilde{\mathcal{F}}\}$.

For each $x\in\bigcup\limits_{F\in\mathcal{F}}N_{F}$, $x\in N_{F_{0}}$ for some $F_{0}\in\mathcal{F}$. It follows form $\up N_{F_{0}}\subseteq U$ that $x\in U$. So $U\in\lozenge(cl(\{x\}))$ and thus $U\in\bigcap\{\lozenge A\mid A\in\widetilde{\mathcal{F}}\}$.

$\mathbf{claim}$ 2: $\bigcap\{\lozenge A\mid A\in\widetilde{\mathcal{F}}\}\subseteq\mathcal{H}$.

Let $V$ be an open set of $X$ with $V\in\bigcap\{\lozenge A\mid A\in\widetilde{\mathcal{F}}\}$. Then for each $x\in\bigcup\limits_{F\in\mathcal{F}}N_{F}$, $V\cap cl(\{x\})\neq\emptyset$. This means that $\bigcup\limits_{F\in\mathcal{F}}N_{F}\subseteq V$. Thus for each $F\in\mathcal{F}$, $\up N_{F}\in\Box V$. Since $Q_{F}\subseteq\up N_{F}$ and $Q_{F}\in F$, $\up N_{F}\in F$. It follows from $\up N_{F}\in\Box V\cap F$ that $\Box V\in\lozenge F$ for each $F\in\mathcal{F}$. Since $\bigcap\{\lozenge F\mid F\in\mathcal{F}\}\subseteq f^{-1}(\mathcal{H})$, $\Box V\in f^{-1}(\mathcal{H})$. Then there are open sets $\{U_{i}\mid i\in I\}\subseteq\mathcal{O}(X)$ such that $\Box V=\bigcup\limits_{i\in I}\Box U_{i}$  and $\bigcup\limits_{i\in I}U_{i}\in\mathcal{H}$. Note that $$V=\bigcup\Box V=\bigcup(\bigcup\limits_{i\in I}\Box U_{i})=\bigcup\limits_{i\in I}(\bigcup\Box U_{i})=\bigcup\limits_{i\in I}U_{i}.$$
So we have $V\in\mathcal{H}$. Hence the claim 2 is proved. By two claims, we can conclude that $X$ is co-consonant.}

Note that the co-consonance of $P_{S}(X)$ does not imply that every compact subset of $X$ is strongly compact. Please see the following example.

\begin{Exam} Let $P=\{\infty\}\cup\mathbb{N}$ and $\mathbb{N}$ be the set of all natural numbers. The order on $P$ is given by:
$$\forall\ x\in P,\ x\leq\infty.$$
Then $X=(P,\upsilon(P))$ is a quasi-polish space (see \cite{WR30} Example 3.2). By Example 3.4(1), $P_{s}(X)$ is co-consonant. It is not difficult to verify that $X$ is a compact space. However the compact subset $P$ is not strongly compact in $X$. So we can assert that $X$ is not co-consonant.
\end{Exam}

\quad The following conclusion is one of the most important conclusions of this paper. This gives a partial answer of question 4.

\begin{Th} Let $X$ be a co-consonant space. If for every natural number $n$,
$$\Sigma(\prod\limits^{n}\mathcal{O}(X))=\prod\limits^{n}\Sigma(\mathcal{O}(X)),$$
then $P_{S}(X)$ is co-consonant.
\end{Th}

\p{Let $\mathcal{F}\in\sigma(\mathcal{O}(P_{S}(X)))$ and $\mathscr{F}\in\mathcal{F}$. Then $\mathscr{F}=\bigcup\limits_{i\in I}\Box U_{i}$ for some family $\{U_{i}\in\mathcal{O}(X)\mid i\in I\}$.
Since $\mathcal{F}$ is Scott open, there is a finite subset $\{U_{k}\mid 1\leq k\leq n\}\subseteq\{U_{i}\mid i\in I\}$ such that $\bigcup\limits_{1\leq k\leq n}\Box U_{k}\in\mathscr{F}$. Define a mapping $\beta_{n}:\Sigma(\prod\limits^{n}(\mathcal{O}(X)))\longrightarrow\Sigma(\mathcal{O}(P_{S}(X)))$ as follows:
$$\beta_{n}(V_{1}, V_{2},\cdot\cdot\cdot,V_{n})=\bigcup\limits^{n}\limits_{i=1}\Box V_{i}.$$
Since $\beta_{n}$ is Scott continuous for each component $V_{i}$, $\beta_{n}$ is continuous. As $\Sigma(\prod\limits^{n}\mathcal{O}(X))=\prod\limits^{n}\Sigma(\mathcal{O}(X))$, $\beta_{n}$ is also continuous form $\prod\limits^{n}\Sigma(\mathcal{O}(X))$ to $\Sigma(\mathcal{O}(P_{S}(X)))$. Then there are finitely Scott open sets $\{\mathcal{H}_{k}\mid 1\leq k\leq n\}\subseteq\sigma(\mathcal{O}(X))$ such that
$$(U_{1},U_{2},\cdot\cdot\cdot, U_{n})\in\mathcal{H}_{1}\times\mathcal{H}_{2}\times\cdot\cdot\cdot\times\mathcal{H}_{n}\subseteq\beta_{n}^{-1}(\mathcal{F}).$$ By the co-consonance of $X$, there is a finite subset $\mathcal{E}_{k}\subseteq\Gamma(X)$ such that $U_{k}\in\bigcap\{\lozenge V\mid V\in\mathcal{E}_{k}\}\subseteq\mathcal{H}_{k}$, for each $1\leq k\leq n$. Let $E_{k}=\bigcap\{\diamond V\mid V\in\mathcal{E}_{k}\}$.
Then $\mathcal{E}=\{E_{k}\mid 1\leq k\leq n\}$ is a finite subset of $\Gamma(P_{S}(X))$.

$\mathbf{claim}$ 1: $\bigcup\limits_{1\leq k\leq n}\Box U_{k}\in\bigcap\{\lozenge E_{k}\mid 1\leq k\leq n \}$.

For each $1\leq k\leq n$, $U_{k}\in\bigcap\{\lozenge V\mid V\in\mathcal{E}_{k}\}$. Then for each $V\in\mathcal{E}_{k}$, $U_{k}\cap V\neq\emptyset$. Take a $x_{k,V}\in U_{k}\cap V$. Let $F_{k}=\up\{x_{k,V}\mid V\in\mathcal{E}_{k}\}$. By the finiteness of $\mathcal{E}_{k}$, $F_{k}$ is compact in $X$. Clearly, $F_{k}\in\Box U_{k}\cap E_{k}$. It follows from $\Box U_{k}\in\lozenge E_{k}$ that $\bigcup\limits_{1\leq k\leq n}\Box U_{k}\in\lozenge E_{k}$. So $\bigcup\limits_{1\leq k\leq n}\Box U_{k}\in\bigcap\{\lozenge E_{k}\mid 1\leq k\leq n \}$ and thus $\mathscr{F}\in\bigcap\{\lozenge E_{k}\mid 1\leq k\leq n \}$.

$\mathbf{claim}$ 2: $\bigcap\{\lozenge E_{k}\mid 1\leq k\leq n \}\subseteq\mathcal{F}$.

For each $\bigcup\limits_{j\in J}\Box W_{j}\in\bigcap\{\lozenge E_{k}\mid 1\leq k\leq n\}$, we have that for each $1\leq k\leq n$, there is a $Q_{k}\in\mathcal{Q}(X)$ such that $Q_{k}\in\bigcup\limits_{j\in J}\Box W_{j}\cap E_{k}$. Then there exists $j_{k}\in J$ satisfying $Q_{k}\in\Box W_{j_{k}}$. As $Q_{k}\in E_{k}$ and $Q_{k}\in\Box W_{j_{k}}$, $W_{j_{k}}\in\bigcap\{\lozenge V\mid V\in\mathcal{E}_{k}\}$. This means that $W_{j_{k}}\in \mathcal{H}_{k}$ for each $1\leq k\leq n$. So we have $$(W_{j_{1}},W_{j_{2}},\cdot\cdot\cdot,W_{j_{n}})\in\mathcal{H}_{1}\times\mathcal{H}_{2}\times\cdot\cdot\cdot\times\mathcal{H}_{n}\subseteq\beta_{n}^{-1}(\mathcal{F}).$$
Whence, $\bigcup\limits_{1\leq k\leq n}\Box W_{j_{k}}\in\mathcal{F}$. Since $\mathcal{F}$ is an upper set in $\mathcal{O}(P_{S}(X))$, $\bigcup\limits_{j\in J}\Box W_{j}\in\mathcal{F}$.
Those two claims imply that $P_{S}(X)$ is co-consonant.}

\quad By Lemma 2.1, Lemma 2.2 and Theorem 4.4, the following conclusion holds immediately.

\begin{Cor} Let $X$ be a co-consonant space. If $X$ is core-compact or $\Sigma(\mathcal{O}(X))$ is first-countable, then $P_{S}(X)$ is co-consonant.
\end{Cor}

\quad The next conclusion demonstrates that co-consonance of original space is necessary for the co-consonance of lower powerspace. Thus we give a positive answer of question 6.

\begin{Th} Let $X$ be a $T_{0}$ space.  If $P_{H}(X)$ is co-consonant, then $X$ is co-consonant.
\end{Th}
\p{Define a mapping $\xi:X\longrightarrow P_{H}(X)$ by $\xi(x)=cl(\{x\})$, for all $x\in X$. Then $\xi$ is a topological embedding. So the mapping $\eta:\mathcal{O}(P_{H}(X))\longrightarrow\mathcal{O}(X)$ is well defined, where $\eta(\mathcal{U})=\xi^{-1}(\mathcal{U})$ for all $\mathcal{U}\in\mathcal{O}(P_{H}(X))$. Clearly, $\eta$ is Scott continuous. Let $\mathcal{F}\in\sigma(\mathcal{O}(X))$ and $U\in\mathcal{F}$. By the definition of mapping $\eta$, we have

\begin{center}$\begin{array}{lll}
\eta^{-1}(\mathcal{F})&=&\{\mathcal{U}\in\mathcal{O}(P_{H}(X))\mid\xi^{-1}(\mathcal{U})\in\mathcal{F})\}\\
&=&\{\bigcup\limits_{i\in I}(\bigcap\limits_{j\in J_{i}}\diamond U_{j})\mid\{x\mid cl(\{x\})\in\bigcup\limits_{i\in I}(\bigcap\limits_{j\in J_{i}}\diamond U_{j})\}\in\mathcal{F}\}\\
&=&\{\bigcup\limits_{i\in I}(\bigcap\limits_{j\in J_{i}}\diamond U_{j})\mid\{x\mid \exists\ i_{0}\in I, x\in\bigcap\limits_{j\in J_{i_{0}}}U_{j}\}\in\mathcal{F}\}\\
&=&\{\bigcup\limits_{i\in I}(\bigcap\limits_{j\in J_{i}}\diamond U_{j})\mid\bigcup\limits_{i\in I}(\bigcap\limits_{j\in J_{i}}U_{j})\in\mathcal{F}\},\\
\end{array}$\end{center}
where $J_{i}$ is finite for each $i\in I$ and $\{U_{j}\mid j\in J_{i}\}\subseteq\mathcal{O}(X)$.
Then $\eta^{-1}(\mathcal{F})\in\sigma(\mathcal{O}(P_{H}(X)))$ and $\diamond U\in\eta^{-1}(\mathcal{F})$.
By the co-consonance of $P_{H}(X)$, there is a finite subset $\mathcal{E}\subseteq\Gamma(P_{H}(X))$ such that $$\diamond U\in\bigcap\{\lozenge A\mid A\in\mathcal{E}\}\subseteq\eta^{-1}(\mathcal{F}).$$
Then for each $ A\in\mathcal{E}$, $\diamond U\cap A\neq\emptyset$. So there exists a nonempty closed $F_{A}\subseteq X$ satisfying $F_{A}\in\diamond U\cap A$. Let $\widetilde{\mathcal{E}}=\{F_{A}\mid A\in\mathcal{E}\}$. Then $\widetilde{\mathcal{E}}$ is finite subset of $\Gamma(X)$. Clearly, $U\in\bigcap\{\lozenge F_{A}\mid A\in\mathcal{E}\}$. Let $V$  be an open set with $V\in\bigcap\{\lozenge F_{A}\mid A\in\mathcal{E}\}$. Then for each $A\in\mathcal{E}$, $V\cap F_{A}\neq\emptyset$. This means that $F_{A}\in\diamond V\cap A$. Hence, $\diamond V\in\bigcap\{\lozenge A\mid A\in\mathcal{E}\}\subseteq\eta^{-1}(\mathcal{F})$. Whence, $V=\eta(\diamond V)\in\mathcal{F}$. This implies that $U\in\bigcap\{\lozenge B\mid B\in\widetilde{\mathcal{E}}\}\subseteq\mathcal{F}$. Therefore, $X$ is co-consonant.}

\quad Before answering question 5, it is necessary to introduce the following concept.

\begin{Def}  A topological space $X$ is called intersection-compatible if for every pair of open sets $U,V\in\mathcal{O}(X)$ and a closed set $W\in\Gamma(X)$, $U\cap W\neq\emptyset$ and $V\cap W\neq\emptyset$ implies $U\cap V\cap W\neq\emptyset$.
\end{Def}

\begin{Rem} If every closed set of a topological space $X$ is irreducible, then $X$ is intersection-compatible. In particular, the Scott spaces of chains are intersection-compatible.
\end{Rem}

\begin{Th} Let $X$ be a co-consonant and intersection-compatible space.  If for every natural number $n$,
$$\Sigma(\prod\limits^{n}\mathcal{O}(X))=\prod\limits^{n}\Sigma(\mathcal{O}(X)),$$
 then $P_{H}(X)$ is co-consonant.
\end{Th}

\p{ Let $\mathcal{F}\in\sigma(\mathcal{O}(P_{H}(X)))$ and $\mathscr{F}\in\mathcal{F}$. Then $\mathscr{F}=\bigcup\limits_{i\in I}(\bigcap\limits_{j\in J_{i}}\diamond U_{j})$, where  $J_{i}$ is finite for each $i\in I$ and $\{U_{j}\mid j\in J_{i}, i\in I\}\subseteq\mathcal{O}(X)$. As $\mathcal{F}$ is Scott open, there exists a finite subset $F_{0}\subseteq I$ such that $\bigcup\limits_{i\in F_{0}}(\bigcap\limits_{j\in J_{i}}\diamond U_{j})\in\mathcal{F}$. For convenience, let $F_{0}=\{1,2,\cdot\cdot\cdot,n_{0}\}$. one can check that $$\bigcup\limits_{i\in F_{0}}(\bigcap\limits_{j\in J_{i}}\diamond U_{j})=\bigcap\limits_{U\in\mathcal{S}_{1}}\diamond U,$$ where $\mathcal{S}_{1}=\{\bigcup\limits_{1\leq k\leq n_{0}}U_{i_{k}}\mid i_{k}\in J_{k}\}$. Let $n=|\mathcal{S}_{1}|$ and $\mathcal{S}_{1}=\{U_{1},U_{2},\cdot\cdot\cdot,U_{n}\}\subseteq\mathcal{O}(X)$. Then $$\bigcup\limits_{i\in F_{0}}(\bigcap\limits_{j\in J_{i}}\diamond U_{j})=\bigcap\limits_{U\in\mathcal{S}_{1}}\diamond U=\bigcap\limits_{1\leq k\leq n}\diamond U_{k},$$  and $\bigcap\limits_{1\leq k\leq n}\diamond U_{k}\in\mathcal{F}$.
Define a mapping $\beta_{n}:\Sigma(\prod\limits^{n}(\mathcal{O}(X)))\longrightarrow\Sigma(\mathcal{O}(P_{H}(X)))$ as follows:
$$\beta_{n}(V_{1}, V_{2},\cdot\cdot\cdot,V_{n})=\bigcap\limits^{n}\limits_{i=1}\diamond V_{i}.$$
Since $\beta_{n}$ is Scott continuous for each component $V_{i}$, $\beta_{n}$ is continuous. As $\Sigma(\prod\limits^{n}\mathcal{O}(X))=\prod\limits^{n}\Sigma(\mathcal{O}(X))$, $\beta_{n}$ is also continuous form $\prod\limits^{n}\Sigma(\mathcal{O}(X))$ to $\Sigma(\mathcal{O}(P_{H}(X)))$. Then there are finitely Scott open sets $\{\mathcal{H}_{k}\mid 1\leq k\leq n\}\subseteq\sigma(\mathcal{O}(X))$ such that
$$(U_{1},U_{2},\cdot\cdot\cdot, U_{n})\in\mathcal{H}_{1}\times\mathcal{H}_{2}\times\cdot\cdot\cdot\times\mathcal{H}_{n}\subseteq\beta_{n}^{-1}(\mathcal{F}).$$ By the co-consonance of $X$, there is a finite subset $\mathcal{F}_{k}\subseteq\Gamma(X)$ such that $U_{k}\in\bigcap\{\lozenge A\mid A\in\mathcal{F}_{k}\}\subseteq\mathcal{H}_{k}$, for each $1\leq k\leq n$. Let $\mathcal{E}_{1}=\{\Box(\bigcup\limits^{n}\limits_{j=1}A_{j})\mid A_{j}\in\mathcal{F}_{j}, 1\leq j\leq n\}$. Clearly, $\mathcal{E}_{1}$ is a finite subset of $\Gamma(P_{H}(X))$.

$\mathbf{claim}$ 1: $\bigcap\limits^{n}\limits_{j=1} \diamond U_{j}\in\bigcap\{\lozenge M\mid M\in\mathcal{E}_{1}\}$.

For each $M\in\mathcal{E}_{1}$, $M=\Box(\bigcup\limits^{n}\limits_{j=1}A_{j})$ and for every $1\leq j\leq n$, $A_{j}\in\mathcal{F}_{j}$.
Since $U_{j}\in\bigcap\{\lozenge A\mid A\in\mathcal{F}_{j}\}$ for all $1\leq j\leq n$ , then $U_{j}\cap A_{j}\neq\emptyset$. Take $x_{j}\in U_{j}\cap A_{j}$. Then $\bigcup\limits^{n}\limits_{j=1}cl(\{x_{j}\})\in(\bigcap\limits^{n}\limits_{j=1} \diamond U_{j})\cap(\Box(\bigcup\limits^{n}\limits_{j=1}A_{j}))$. So $\bigcap\limits^{n}\limits_{j=1} \diamond U_{j}\in\bigcap\{\lozenge M\mid M\in\mathcal{F}\}$ and thus $\mathscr{F}\in\bigcap\{\lozenge M\mid M\in\mathcal{E}_{1}\}$.

$\mathbf{claim}$ 2: $\bigcap\{\lozenge M\mid M\in\mathcal{E}_{1}\}\subseteq\mathcal{F}$.

For each $\mathscr{A}=\bigcup\limits_{s\in S}(\bigcap\limits_{t\in T_{s}}\diamond V_{t})\in\bigcap\{\lozenge M\mid M\in\mathcal{E}_{1}\}$, where $T_{s}$ is finite for each $s\in S$ and $\{V_{t}\mid t\in T_{s}\}\subseteq\mathcal{O}(X)$. Similarly, since $\bigcap\{\lozenge M\mid M\in\mathcal{E}_{1}\}$ is Scott open in $\mathcal{O}(P_{H}(X))$, there are finite open sets $V_{1}, V_{2},\cdot\cdot\cdot V_{m}$ of $X$ such that $\bigcap\limits^{m}\limits_{i=1}\diamond V_{i}\subseteq \mathscr{A}$ and $\bigcap\limits^{m}\limits_{i=1}\diamond V_{i}\in\bigcap\{\lozenge M\mid M\in\mathcal{E}_{1}\}$. Suppose that there is a $V_{r}$ ($1\leq r
\leq m$) such that $V_{r}\not\in\bigcap\{\lozenge A\mid A\in\mathcal{F}_{k}\}$ for each $1\leq k\leq n$. Then for each $1\leq k\leq n$, there exists $B_{k}\in\mathcal{F}_{k}$ satisfying $B_{k}\cap V_{r}=\emptyset$. Let $M_{0}=\Box(\bigcup\limits^{n}\limits_{k=1}B_{k})$. Clearly, $M_{0}\in\mathcal{E}_{1}$ and $\bigcap\limits^{m}\limits_{i=1}\diamond V_{i}\in\lozenge M_{0}$. So there is a $F\in\Gamma(X)$ such that $F\subseteq\bigcup\limits^{n}\limits_{k=1}B_{k}$ and $F\cap V_{r}\neq\emptyset$. This contradicts with $B_{k}\cap V_{r}=\emptyset$, for each $1\leq k\leq n$. Then we can see that for each $1\leq i\leq m$, there is a $1\leq k\leq n$ such that $V_{i}\in\bigcap\{\lozenge A\mid A\in\mathcal{F}_{k}\}$. For each $1\leq k\leq n$, let $G_{k}=U_{k}\cap\{V_{i}\mid 1\leq i\leq m,V_{i}\in\bigcap\{\lozenge A\mid A\in\mathcal{F}_{k}\}\}$. Since $X$ is a intersection-compatible space, $G_{k}\in \bigcap\{\lozenge A\mid A\in\mathcal{F}_{k}\}$ for each $1\leq k\leq n$ and $\bigcap\limits^{n}\limits_{k=1}\diamond G_{k}\subseteq\bigcap\limits^{m}\limits_{i=1}\diamond V_{i}$.
Then we have $$(G_{1},G_{2},\cdot\cdot\cdot,G_{n})\in\mathcal{H}_{1}\times\mathcal{H}_{2}\times\cdot\cdot\cdot\times\mathcal{H}_{n}\subseteq\beta_{n}^{-1}(\mathcal{F}).$$  Whence, $\bigcap\limits^{n}\limits_{k=1}\diamond G_{k}\in\mathcal{F}$. Since $\mathcal{F}$ is an upper set, it follows from $\bigcap\limits^{n}\limits_{k=1}\diamond G_{k}\subseteq\bigcap\limits^{m}\limits_{i=1}\diamond V_{i}\subseteq\mathscr{A}$ that $\mathscr{A}\in\mathcal{F}$. So the claim 2 is proved.

These two claims are enough for the proof.}

\end{document}